# Inequalities, chance and success in sport competitions: simulations vs empirical data

Pawel Sobkowicz[1*], Robert H Frank[2], Alessio E Biondo[3],

Alessandro Pluchino[4,5] and Andrea Rapisarda[4,5,6]

[1]National Centre for Nuclear Research, A Soltana 7, 05-400 Otwock, Poland.

[2]Cornell SC Johnson College of Business, Sage Hall Cornell University, NY 14853-6201 Ithaca, USA.

[3]Department of Economics and Business, University of Catania, Palazzo delle Scienze, Corso Italia 55, 95129 Catania, Italy.

[4]Department of Physics and Astronomy "Ettore Majorana", University of Catania, Via S. Sofia, 64, bldg 6, 95123 Catania, Italy.

[5]Instituto Nazionale di Fisica Nucleare, Via S. Sofia, 64, 95123 Catania, Italy.

[6]Complexity Science Hub Vienna, Josefstadter Strasse 39, 1080 Vienna, Austria.

Acknowledgements

A.P. and A.R. acknowledge financial support by the project "Linea di intervento 2" of the Department of Physics and Astronomy "Ettore Majorana" of the University of Catania.

Corresponding author:

Pawel Sobkowicz, National Centre for Nuclear Research, A Soltana 7, 05-400 Otwock, Poland, pawel.sobkowicz@ncbj.gov.pl, tel +48 606 267 122

# Abstract

We present a new way of estimation of the role of chance in achieving success, by comparing the empirical data from 100-meter dash competitions (one of the sports disciplines with the most stringent controls of external randomness), with the results of an agent-based computer model, which assumes that success depends jointly on the intrinsic talent of the agent and on unpredictable luck. We find a small, but non-zero contribution of random luck to the performance of the best sprinters, which may serve as a lower bound for the randomness role in other, less stringently controlled competitive domains. Additionally we discuss the perception of the payoff differences among the top participants, and the role of random luck in the resulting inequality.

## 1. Introduction

Inequalities are present in many domains of social life. In many cases, they are considered unjust and harmful, and societies take active measure to minimize them. In other cases, inequalities are not only accepted but cherished and admired. Such a positive perception of inequalities is particularly evident in sports and arts. The differences in acting quality are rather elusive to measure, yet, in contrast, the differences in the pay of the actors are quite huge. The leading ones can easily demand pay rates in excess of 10-20 million US$, plus additional benefits, in many cases exceeding 50 million US$ (Wikipedia WEB page a, 2018), while the majority of actors, perhaps equally talented ones, have rather meagre incomes. Yet, there is very little outrage about this inequality: on the contrary, the presence of a highly paid (overpaid?) star is one of the best ways to ensure a movie box office success. Of course, superstars do not guarantee film success. However, quite often they act as magnets, attracting the sufficient number of moviegoers to make the film profitable, even after their huge salaries are taken into account. The viewers accept and support the inequality with their feet and wallets. Whether the star effect is real, confirmed by the correlation of the box office results and the presence and pay of the stars, may be debated (Wallace et al., 1993; De Vany and Walls, 1999, 2004). But it is certainly perceived as true by the producers, as evidenced indirectly by the dominant position of star actors in promotion and advertising. Today, the acceptance of the exorbitant income of celebrities (actors, performers, sportsmen) may also be measured more direct measures, for example by the number of people who 'follow' them on social media. Sites tracking such popularity for either Facebook and Twitter (Wikipedia WEB page b, 2018; FanPage WEB page, 2018) or Instagram (Statista WEB page, 2018) clearly show that the most successful entertainers have tens of millions of followers. To quote a CNBC article (Giuliano, 2015) *'Massive bank account and massive social media following? Check and check.'* The same acceptance of inequality is seen in sports. The emotional identification and admiration for the champions is so universal that it has even been the basis of an accusation of resembling fascism (Tännsjö, 1998).

The differences of performance are the very essence of sports competitions, deciding winners and losers, thus making the show interesting. Sports define performance via quantitative measures, identifying success with deserving, earned qualities of well-developed talent, effort and persistence in training. This focus on 'objective' differentiation, supposedly provided by numerical measures is present even in disciplines where the performance is only qualitatively evaluated by human judges, for example, figure skating or gymnastics[1].

As in arts, since the late 20th century, the rewards for winning in sports have become very high and steep. Top soccer players not only earn huge salaries from their clubs; they also reap the indirect benefits via even higher advertising contracts. One can argue that in this case, their higher performance is more visible and measurable than the actors' talent. However, if we consider, for example, the number of goals scored, it is distributed much less extravagantly than the corresponding players' incomes. Similar 'winner worship' is present in many other sports. Who remembers the runner who has been fourth in the final 100 meter race in the Rio Olympic Games in 2016? Probably quite a few people would recall that the winner was Usain Bolt, acclaimed as 'the fastest man on Earth', but who took the fourth place? Yet there was only 0.12 second difference in their results, about 1%[2]. The organizers of track and field events – just like film producers – try to attract stars to ensure larger numbers of spectators. A part of this approach is to create very steep financial awards or even direct 'appearance fees', not connected to actual performance. The prizes range from moderate (for example in the IAAF Diamond League Champion wins US$ 50,000, the second place awards 20,000, the third 10,000) to quite big (the Wimbledon tennis championships award singles winners of both sexes 2,250,000 pounds, finalists 1,125,000 pounds and semi-finalists 562,000 pounds). The Wimbledon prizes have more than doubled since 2010 (Wimbledon WEB page, 2018). This exponential disparity in financial rewards hardly corresponds to an exponential disparity in the talent of the players who advance to the final levels. Yet, there are very few complaints and calls for the abolition of such blatant inequality.

The role of personal talents, dedication and effort put into preparations versus chance or luck in sports performance has been a subject of several studies. To mention a few examples, Deng et al. (2012) have studied the universality of scores and prizes in a large number of disciplines. Radicchi (2012) paper focused on the progress of the winning results in several Olympic disciplines (including the 100 meter sprint) and the ultimate limits of performance, including the data on the statistical distribution of the winners' results. Yucesoy and Barabási (2016) have analysed the disparities in tennis performance and rewards. All these works were motivated by the fact that sports are devised to provide a highly competitive environment, coupled with a setting that strengthens the objective measure of the performance. This clarity is absent in many other fields, where not only we do not have the direct knowledge of the underlying talents, but also the resulting performance might not be comparable or even measurable.

In contrast to these examples of the acceptance of extreme inequalities in sports and entertainment, there are many domains in which inequality is highly contested and considered socially harmful. Of course, the most visible of these is the inequality of wealth and income. There are many voices calling for active measures to alleviate wealth inequalities through redistribution. The discussions related to factors determining individual careers, promotion and income often focus on the role of personal endowments (especially intelligence, family environment, race and social status) and on the role of chance events, unrelated to individual characteristics.

---

[1] Similar 'quantitative objectivity fetish' is present in science, with its growing reliance of research evaluation on bibliometric or altmetric parametrization.

[2] Fourth place was taken by Yohan Blake from Jamaica, silver medallist from the previous Olympic Games in London. It is worth noting that the personal best for Blake, as recorded by IAAF is 9.69 s, measured with headwind of 0.1 m/s, while the best for Bolt is 9.58, recorded with tailwind of 0.9 m/s.

It is not our purpose here to provide an in-depth discussion of the philosophical, psychological and social analyses of the origins, validity and practical measures related to equality and forms of its implementation, such as the 'equality of opportunity' or 'equality of outcome'. The literature on the topic is extremely rich, covering issues of fundamental motivation of equality (Rawls, 1999; Anderson, 1999); studies of differences in categorizing earned vs non-earned benefits from a psychological perspective and dependence of reward allocation on luck versus effort attribution (Wittig et al., 1981; Thompson and Prendergast, 2013; Jones, 2015); differences in perception and acceptance or rejection of social justice measures designed to promote equality (Joseph, 1980; Oppenheim, 1980; Pojman, 1995; Arneson, 2000; Bénabou and Tirole, 2006; Almas et al., 2010; Kuppens et al., 2018).

The wealth, and especially the income inequalities, are often defended on the basis of the 'just desert' notion: i.e. the idea that high pay and accumulated wealth result, at least in large part, from hard work and personal abilities, and therefore are rightfully earned. It is worth noting that such an explanation is in conflict with the evidence that differences in personal characteristics and efforts are too small to justify, by itself, the span of income distribution. Actually, intelligence variation - as defined, for example, by the normal IQ scale, with most people's scores falling between 70 and 130 - is rather limited compared to the enormous differences in income and wealth, which span orders of magnitude. The same holds for the efforts, quantified, for example, by the working hours, which are also limited and, typically, normally distributed. In contrast to the limited variation of the variables used to explain inequalities, the distribution of earnings and wealth is much broader, power-law like, as noted first by Pareto (1896)..

A possible key for understanding such a macroscopic discrepancy between inputs (efforts and abilities) and outputs (incomes and capital) lies in the structure and in the complexity of our globally networked socio-economic system, full of feedback mechanisms and winner-take-all domains. In this highly non-linear context, the use of a simple linear paradigm connecting intellectual capacity or productivity efforts with the scale invariant distributed wealth score does appear rather naive. In reality, it frequently happens that small advantage/disadvantage in IQ or small differences in efforts could lead to large increase/decrease in the resulting income, since the latter may be strongly influenced by cumulative effects induced by the multiplicative dynamics of the system. DiPrete and Eirich (2006) have reviewed the work on the cumulative advantage role in growth of inequalities. They have described many mechanisms leading to accumulation of advantage. For example they have compared the 'superstar' model of Rosen (Rosen, 1981; Bowbrick et al., 1983; Rosen, 1983), who argued that small differences in innate talent can lead to large differences in earnings, with the approach of Merton (Merton et al., 1968; Merton, 1988), who stressed the role of external circumstances and resources, which might lead to situation in which lower innate talent can result in higher output and awards. In the context of intelligence based advantages, the cumulative advantage may be present at even deeper level, namely genotype-environment correlation, where small genetic differences are strengthened '*when children select, modify and create environments correlated with their genetic propensities*' (Plomin and Deary, 2015).

Coming back to sports – where the income advantages are typically expected to be due to differences in physical abilities – one could try to explain the nonlinear effects in performance by the training. Indeed, it is true that the most talented sportsmen benefit from working with the best coaches and submit themselves to the most rigorous training regime. But this may realistically explain differences between amateurs and professionals. The level of coaching and the effort put into workouts by, say, top-5 performers of a discipline, is very similar. Yet, the difference between rewards of the winner and those who end the competitions 'just below the podium' are, in some cases, staggering.

Our interest in sports is motivated by the fact, that at least in some disciplines, the performance of individual participants is accurately and quantitatively measured. This stands in contrast with other social domains where devising such simple measures is rather difficult or even impossible. Moreover, sport competitions are designed to provide equal conditions to all participants, to the maximum possible degree. Artificial 'boosters', such as doping, are grounds for disqualification. In many disciplines reputation or previous results might help a little (for example by allowing the past champions to join competition at later stages, skipping the preliminary eliminations, or seeding at tennis tournaments), but eventually even the most decorated sportsmen must face their rivals in a fair competition.

These conditions make sport competitions well suited to try novel research tools, such as Agent Based Models, in an effort to understand the role of individual abilities and external circumstances in determination of performance and in achieving success. Denrell and Liu (2012) have proposed a model in which the performance in a series of 50 games results from a combination of skill and noise. They came to the conclusion that it is not obvious that higher performance indicates higher skill, especially when the noise is high. Baek et al. (2013) have considered a different setting of a simple knock-out tournament and studied the influence of 'imperfect' resolution in individual matches, that is the probability that a less skilled person can win against a more talented one. In addition, the model considered the distributions of prize money.

### 1.1 Motivation

The provision of an explanation for any social ranking is always intriguing and challenging. Specially for income differences, strong temptations arise to justify inequalities by differences in talents and abilities, which is a very contentious issue. A founded criticism of meritocracy is provided by Frank (2016), which contains a thorough analysis of the role of luck in life, challenging the validity of meritocratic paradigm. In his book, Frank considered a single, global competition, in which agents compared performance, defined as a linear combination of their innate talent and a certain amount of random luck. In other words, Frank has been looking for the 'perfect ranking' in presence of noise. Results have shown that, even if the luck part is rather small, it is quite improbable that the most talented agent would end up as the ultimate winner. In fact, Frank (2016), states that in a thousand repetitions of the contest simulations '*only a small minority of winners would have higher combined skill and effort levels than **all other** contestants*' (p. 11). This result, especially when applied to the winner-take-all domains, would lead to huge rewards achieved largely thanks to luck. It is taken as a proof that meritocracy, defined by equating the performance and the ranking based on it with merit (talent, effort or skill), is largely unjust, as many deserving candidates fail to achieve success due to random misfortune. We note, however, that Frank repeats numerous times in his book that, while the ultimate winners are most likely not the ones with the highest talent, '*no one could win without being both talented and hard-working*' (p.11) – and lucky, at the same time. In other words, hard work and talent are necessary to succeed, but may simply be insufficient. The role of luck, chance and circumstances beyond personal control needs to be recognized in shaping the rewards system in society.

This need is well reflected by a recent book by Barabási (2018), which summarizes research on the relationship between achieved success and factors such as performance, effort, persistence, networking and recognition, and individual histories. BarabÃ¡si rightly notes that performance must be empowered by opportunity to lead to a success (p. 27). Moreover, in circumstances where performance can not be measured. success is driven by networking and recognition (p. 62). Still, we may ask about domains where every effort has been taken to allow accurate measurements of performance and to make the 'playing field' equal to everyone. Would success in such fields reflect directly the distribution of talent, effort an persistence, or would other factors, including random luck, play a role?

The second inspiration for our work is the agent-based model presented in Pluchino et al. (2018). It is based on quite different details than the previous one, focusing on simulated lifetime accumulation of wealth due to random occurrence of lucky or unlucky events. The main assumptions of this model are the normal distribution of talent among the agents and the multiplicative dynamics of income/success (Matthew effect). The latter effect, as discussed in the introduction, is due to the highly non-linear nature of the global socio-economic system, which not only induces a macroscopic discrepancy between inputs (talent, effort) and outputs (success), but makes our life very sensitive to small random perturbations, that we interpret as good luck or bad luck. In the model, all agents start with the same amount of capital and are periodically exposed to a variable number of events: a lucky event could lead to the agent doubling its capital/success - but only with a probability equal to her/his talent (limited between 0 and 1). On the other hand, an unlucky event resulted in halving of the capital, regardless of the talent. The final capital distribution was roughly power-law like, with an exponent of $-1.27$ (consistent with the well known Pareto law). In other words, a large number of agents ended the simulation with low values of capital and only a few became rich and very rich. On the other hand, the richest agents turned out not to be the most talented ones, but the ones that were consistently lucky and only moderately talented.

Both the previous models have used relatively simple simulations focusing on particular outcomes. Despite the differences in the considered cases, they reach the same general conclusion: very often the winning agent is not the most talented but the luckiest. At the same time, from both the two models, it also results that a certain degree of talent is necessary to reach the top of success, but it is not sufficient since talent needs opportunities (i.e. good luck) to be effective. Thus, if from an individual point of view talent is fundamental, for it gives more chances to grasp random opportunities, from a social point of view, an individual who is both very lucky and very talented is much rarer to find than an individual who is very lucky and just moderately talented, because very talented people are a very small minority. In other words, the joint probability of high talent and luck is obviously smaller than the one of just luck. Such findings stand as drawbacks of the meritocratic paradigm, which is further challenged by evidences from all contexts in which the personal talent is difficult to be evaluated *a-priori* and past performances of people are used as a proxy of their ability:

in such situations, the risk is to reward people that, at the end of the day, were merely lucky. Thus, a common - albeit myopic - form of meritocracy could be defined 'naïve'. A true meritocratic approach should take into account the combination of multiplicative feedback mechanisms and randomness instead, and design sound policies (as suggested by Frank and Pluchino, Biondo and Rapisarda) to allow the most talented people to attain the top of the social ladder. Moreover, many social benefits would also emerge for the community as a whole, because top roles occupied by mediocre – but lucky – individuals may result in undesirable effects on society.

The research question of the present paper is: what happens if one considers much more specific winner-take-all contexts? In particular, we focus on the environments, where repeated competitions lead to a gradual winnowing of the initial participant pool. The examples are sports or musical competitions, exams deciding the progress in the education system, or periodic evaluation of employee performance which might lead to a promotion or not. A common characteristic is that at each step a smaller number of participants advances to the next (upper) level of the ranking, while the losers exit forever. In such a configuration, talent intuitively appears much more important in order to reach the top of the ranking. Then, if this were true, meritocracy would be automatically ensured by the strong selection realised in the first rounds. There are two more issues to be considered. First, the randomness could still play some role: how often the most talented people end up in the top of the ranking if the random luck influence is present? How vital could be the role of chance in choosing the winner among a small group of very talented finalists? After all, it is evident that a tennis ball falling within or outside the tennis court after hitting the net can strongly affect the match. Therefore, answering to these questions is interesting because, also in the case in which the winner and the other agents near the top of the ranking would have 'almost the same' very high value of talent, the role of randomness can be not negligible in determining the final winner. Secondly, the highly asymmetric scale of rewards, typical of these winner-take-all contexts, is hardly justifiable once it is clear that the difference in talent between two finalists are extremely smaller than the one of their awards received after the match.

Stimulated by this scenario and prompted by the methodology developed in the works of Frank and Pluchino *et al.*, we have decided to build a new model that would combine the most significant features of the previous approaches in order to explore, through large-scale simulations, the dynamics emerging from the competition between agents in a generic but modifiable winner-take-all context, involving repeated rounds in a tournament style. The specific model framework (number of tournament rounds, size of the competing groups) was chosen for simplicity, but also for its similarity to certain sport disciplines. This has allowed to test the approach by comparing its results to empirical data, without losing in generality.

Our intention in creating the model is to compare its results with a suitably similar real world situation. We have chosen 100 meter dash Olympic results because of the availability of detailed quantitative data and stringent controls of the external circumstances, which are expected to minimize the role of chance. We note here that while the role of luck in sports has been previously debated, from the perspective of specific sports and also from the point of view of justification of rewards and their deservedness (Bailey, 2007; Simon, 2007; Johnson et al., 2008; Aicinena, 2013; Schweiger, 2014; Morris, 2015; Loland, 2016), most of these analyses are based on anecdotal evidences and philosophical considerations. To our knowledge there were few attempts to analyse the impact of chance quantitatively, and the adoption of an agent-based models generating data to effectively match empirical statistics is a novel contribution to the subject.

We note here, that while the 100 meter sprint is widely recognized as one of the most important track and field disciplines, it is not particularly lucrative. Usain Bolt net worth is estimated at 60 million US$, followed by Carl Lewis (20 million), Tyson Gay (12 million), Michael Johnson (7 million) and Asafa Powell (6.5 million) (Celebrity WEB page, 2018). None of them were even close to being among the top 100 richest athletes, which required over a 100 million US$ net worth (TheRichest WEB page, 2018). Nevertheless, the gap between Bolt and his top competitors is staggering.

Before we turn to details of the model, we would like to clarify the context in which some words would be used in this paper. For the sake of clarity, we assign the notion of '*talent*' to innate personal traits relevant to the issue (which may be physical prowess and stamina in sports, looks and acting talent in film industry or intelligence, knowledge and dedication in career development); the notion of '*social circumstances*' to the effects of the environment a person has been brought up in (family, societal stratum, educational opportunities available etc.); and finally '*luck*' to denote the random component (e.g. variation in mental or physical disposition influencing specific result, presence of external factors acting in some way such as headwind/tailwind in a competition[3]). Using this language, one may formulate a general research question: what are the relative roles of 'talent', 'social circumstances' and 'luck' in achieving high rank and rewards associated with it.

In what follows we have decided to limit our analysis by omitting the effects of social circumstances. As the approach is based on simplified computer modelling, it would be impossible to include all the complexity of the social systems which favour or limit people's ability to pursue a career, to use their talents. We focus therefore on situations and groups of people which are able to compete for a given position or reward. This, of course, excludes systemic inequalities or biases that prohibit parts of societies or individuals from participation – a very important topic, indeed – from our work.

Even with this limitation, the variety of the ways people actually compete for positions is too big for a single, universal model. For this reason, in this paper we consider a subset of the 'talent vs luck' issues, namely **situations when the ranking depends on a series of comparisons of results within the same setting, divided into separate stages in which ever smaller number of agents are allowed to participate**. Additionally, we focus mainly on the resulting ranking, not on the associated rewards and their scale or justification. These comparisons might take the form of sports competitions, series of exams deciding the progress in the education system, or periodic evaluation of employee performance which might lead to promotion or not. A common characteristic is that with each step a smaller number of participants advances to the next (upper) level of the ranking. We stress the fact that outside specially designed laboratory experiments it is quite impossible to reliably decouple the effects of talent and luck on individual performance. In other words, when the comparisons are used to rank people, to divide them into 'winners' and 'losers', they rely on the observable characteristics: measurable results in athletics, exam scores and teachers' grades, sales figures or supervisors' opinions – which are then treated as a proxy for the actual capabilities of a person. This simplification, while largely unavoidable, should remain on our minds when we make decisions about other people based on their ranking.

---

[3] The role of the wind in the 100 meter competition may seem questionable: the runners of a specific race compete in practically the same conditions. But even here it may be present, if the conditions are different between various races at a given elimination stage and if the rules state that in addition to the first 2–3 runners in each race, a few 'lucky losers' with best times also qualify. There are other disciplines where the wind can play a randomizing role, for example long jump/triple jump, javelin and discus throws, cycling and ski races in which the participants start at different times. Probably the most affected discipline is ski jumping, where recently wind measurements have been explicitly included in the algorithm calculating the points from the raw jump distance.

## 2. Model description
### 2.1 Tournament ranking process
The model is intended to apply to cases when people undergo repeated comparative tests of similar nature (i.e. testing the same characteristics), and where certain ratio of the participants 'passes' to enter the next stage, to compete among themselves. We have simplified this repeated tournament in the following way. The model starts with a relatively large number of agents (1 million). They are divided into groups of ten agents, competing with each other. From each group the agent with the highest performance advances to the next round. It takes just 6 rounds to select the single winner. In the following we will also use the notion of 'finalists' to describe the ten winners of round 5 of the tournament and semi-finalists for the 100 winners of round 4.

The next element of the model is the calculation of performance. Each agent $j$ is characterised by an unchanging value of its 'talent' $T_j$. This value is randomly drawn from a probability distribution, which is shared by all agents (more details on the distributions used are in the next subsection. At each tournament round $t$ in which the agent $j$ participates, a random value of 'luck' $L_j(t)$ is assigned to it, drawn from a uniform distribution limited between 0 and 1. This luck may originate from external circumstances impacting the agent's performance in this concrete competition, from other internal (and changeable) characteristics which might also influence the performance, etc. The agent's performance in round $t$, $P_j(t)$, is then calculated as a linear combination:

$$P_j(t) = aT_j + (1-a)L_j(t) \tag{1}$$

The parameter $a$ describes the importance of talent in the competitions, conversely, $(1-a)$ describes the importance of chance (luck). For $a = 1$ we have 'perfect' competitions, in which the most talented agent in the competing group always wins. For $a < 1$, the influence of luck may result in a less talented, but luckier, agent winning the competition in its group. If they are 'consistently lucky', even less talented agents may end up in the advanced tournament rounds. In other words, the winners are not always the most 'deserving' (i.e. talented) agents, many of whom lose at some stage to the more lucky competitors. To obtain statistically stable results we have run the entire tournament 1000 times for each configuration.

### 2.2 Assumptions for the talent distribution
In our model we have assumed that the agent's talent is limited between 0 and 1. For simplicity, we have assumed that the talent distributions are symmetric in this range, centered around 0.5. With this assumption, the key factor turns to be the width of the talent distribution. Two forms of such distributions were used: the uniform distribution and a truncated Gaussian one. For the Gaussian distributions we have used two values of the standard deviation $\sigma_T$: 0.1 and 0.2. The shapes of the three distributions are shown in Figure 1. The most important difference (in the context of the tournament) is the relative number of highly talented agents. For the narrow Gaussian distribution ($\sigma_T = 0.1$) there are very few agents with large talent values (in the population of 1 million agents, we expect approximately 31 agents with talent above 0.9 and just 3 with talent above 0.95). For the distribution with $\sigma_T = 0.2$ the talented agents are relatively numerous (approximately 16750 with talent above 0.9, 6090 above 0.95, and almost 950 with talent above 0.99). These numbers are especially important when we compare them with the size of the available topmost positions (100 positions in competition round 5, semi-finals, and 10 positions in round 6, the finals). For relatively broad talent distributions there are many more extremely talented candidates than the number of places at the top of the hierarchy.

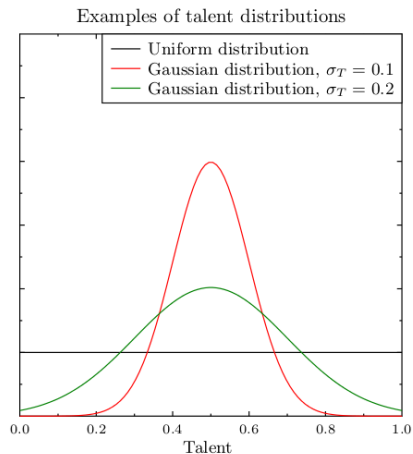

Figure 1: Talent distributions used in the simulations. All distributions are bounded between 0 and 1 and share the same average equal to 0.5. The crucial difference for our analysis is in the number of agents with the highest talent. Out of 1 million agents in total, there are only 3 agents with talent greater than 0.95 for $\sigma_T = 0.1$. At the same time, for $\sigma_T = 0.2$ the number is 6090.

There are therefore only two parameters describing the simulation configurations: the width $\sigma_T$ of the talent distribution (the uniform distribution can be considered as a Gaussian with $\sigma_T = \infty$), and $a$, defining the importance of talent in calculating the performance.

## 3. Model results
### 3.1 Performance as function of talent/luck balance

Figure 2 presents the average performance and talent values for the winners of the first five competition rounds (i.e. values for the agents who advanced to the next round) for the three adopted talent distributions as functions of $a$. The performance (especially for the early rounds) depends on $a$ non-monotonically. For very small $a \approx 0.1$, there is almost no improvement of average performance with each step in the competition. The explanation is quite obvious: at each stage the winners are the luckiest agents, with talent playing a minor role. For $a = 0$ (pure luck competitions) the expected average value is independent of the competition stage and equal to 0.9.

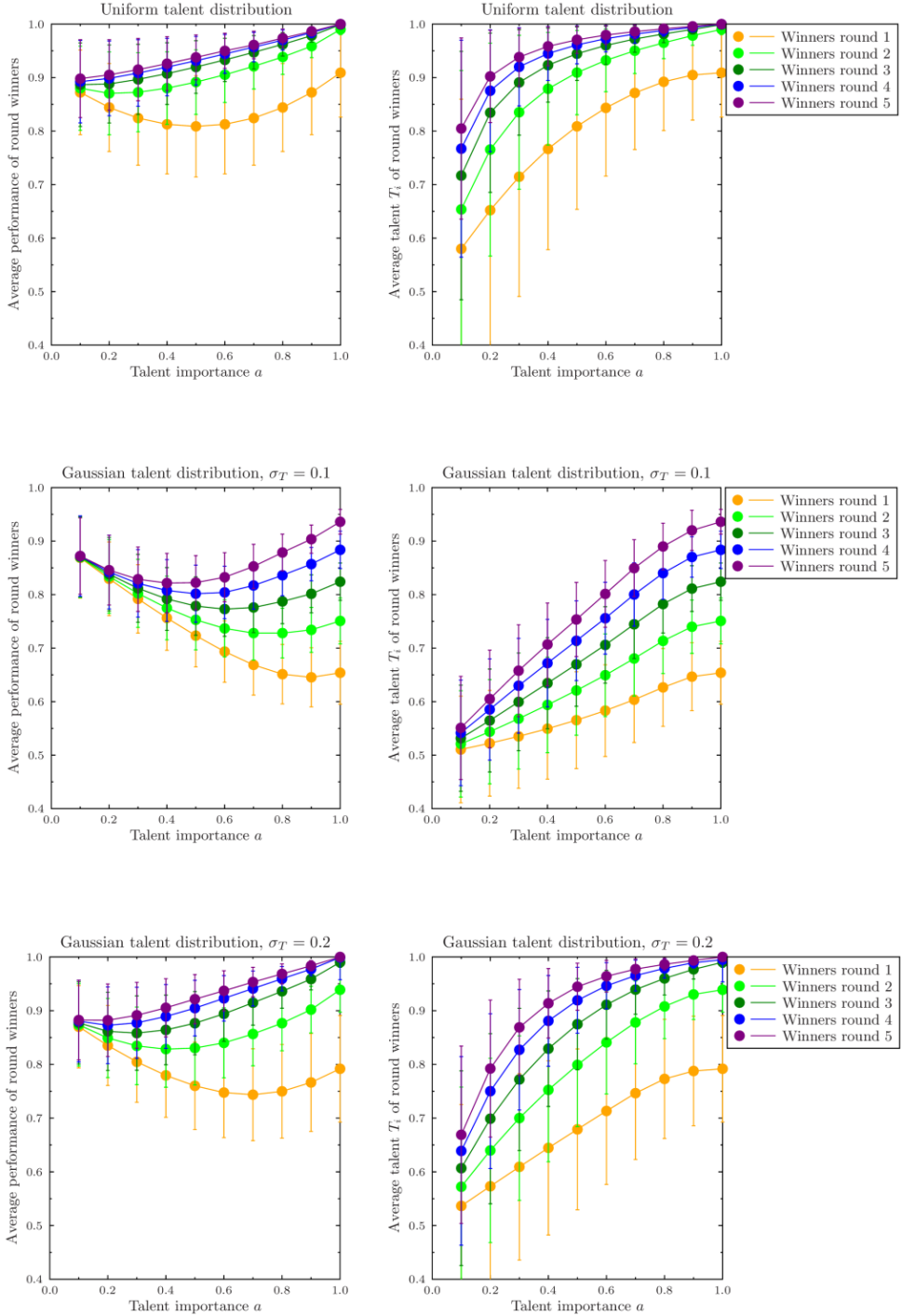

Figure 2: Average performance and talent of round 1–5 competition winners. Performance and talent are presented as functions of the talent importance parameter $a$ for the uniform talent distribution (top panel) and for the two Gaussian distributions with, respectively, $\sigma_T = 0.2$ (middle panel) and $\sigma_T = 0.1$ (bottom panel). Error bars denote the standard deviation averages in each group.

Increasing $a$ leads to a growing separation of the average performances between the competition turns. As talent plays increasingly important role in selecting the winners for larger $a$ values, each subsequent turn starts with a higher average talent value (shown in the right panels). This leads to improved performance in the next round, as the luck component remains the same on the average between the turns.

The average values contain important, but limited information about the winners of the competition rounds. More information is provided by detailed distributions of performance among the winners at various stages. Figure 3 presents two examples of such distributions, for the broad talent distribution ($\sigma_T = 0.2$) and for two values of the importance of talent: $a = 0.2$ and $a = 0.5$. We draw the attention to the fact that both cases would be considered, in real life, to represent very high luck contribution to success.

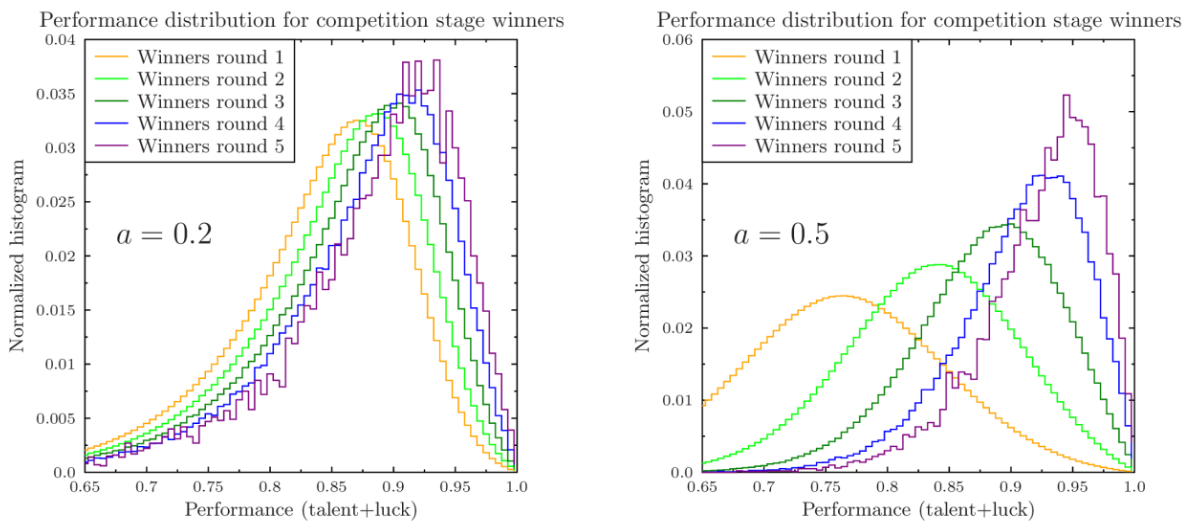

Figure 3: Performance distribution among the winners of five competition stages. Both panels use the same width of the talent distribution $\sigma_T = 0.2$. Left panel corresponds to very low importance of talent ($a = 0.2$), while the right panel corresponds to a case in which half of the performance is due to luck ($a = 0.5$).

While in the very low talent importance case of $a = 0.2$ the shape of the performance distribution among the winners of each round changes but a little, for the moderate value of $a = 0.5$ the winners of the later rounds have achieved much higher performance – and the distribution shape changes significantly. To understand this difference we should look into the distribution of talent among the winners.

### 3.2 Talent distribution among winners

We have already noted that for large enough values of $a$ agents advancing to later rounds of the competition have, on the average, larger talent values. If the talent distribution is broad enough (for example $\sigma_T \geq 0.2$) the average talent of the winners of rounds 4 or 5 is above 0.9 – very high indeed (see Figure 2). The fact that this average is smaller for the narrow talent distributions is simply due to the fact that there are too few high talent agents available. Let us now return to the cases depicted in Figure 3 by comparing them with the corresponding distributions of talent among the winners of each round presented in Figure 4.

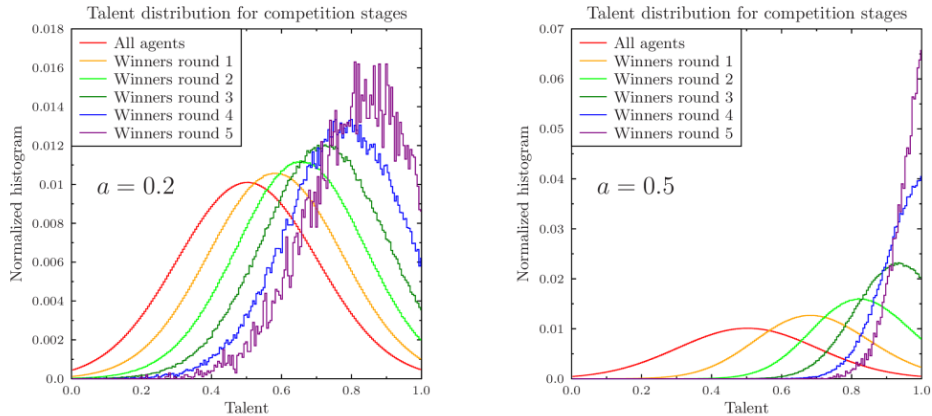

Figure 4: Talent distribution among the winners of competition stages 1–5, for a broad talent distribution. The overall initial talent distribution is shown for reference (red line, $\sigma_T = 0.2$). The parameter values are the same as in Figure 2. Increasing the talent contribution $a$ radically shifts the distributions of talent among the winners of consecutive stages towards higher values. Comparing these results with those of Figure 2, we note that for the agents with the highest talent, chance takes typically the role of bad luck, decreasing performance. For these agents one can expect that $L_j(t) < T_j$, resulting in performance being smaller than talent $P_j(t) < T_j$. For this reason, among the finalists, talent is distributed closer to the maximum value of 1 than the performance.

Even for very low importance of talent ($a = 0.2$), each competition round 'selects' agents with slightly higher talent values. What happens for the moderate value of $a = 0.5$ is stunning: practically no agents of talent below 0.5 could win round 3 of the tournament, this limit increased to about 0.7 for round 4 and 0.8 for round 5. In other words: practically all the winners of stage 5 had talent values over 0.8, despite the fact that at each round luck contributed 50% to the performance. For those agents with the highest talent, the variability of luck is actually more likely to decrease the performance, than to increase it (which can be seen by comparing the right panels of Figures 3 and 4. Figure 4 also answers the question whether, in this case, we can use the performance (as measured for example by winning late competition rounds) as a reliable proxy for the 'hidden, unmeasurable talent. Actually, even when luck role is relatively high, (the example shown uses $a = 0.5$), if the process involves several elimination steps (tournament style, as opposed to a single stage selection) and the distribution of talent is large enough, the winners of the late stages of the tournament are almost certainly very talented.

The shift of the distributions of performance and talent to higher values as the role of luck diminishes is visible as well for the narrow talent distributions. In this case, however the small number of agents with high talent determines a change in the shape of the distributions (see Figures 5 and 6) and one needs to increase the parameter $a$ in order to find very talented winners in the late stages of the tournament.

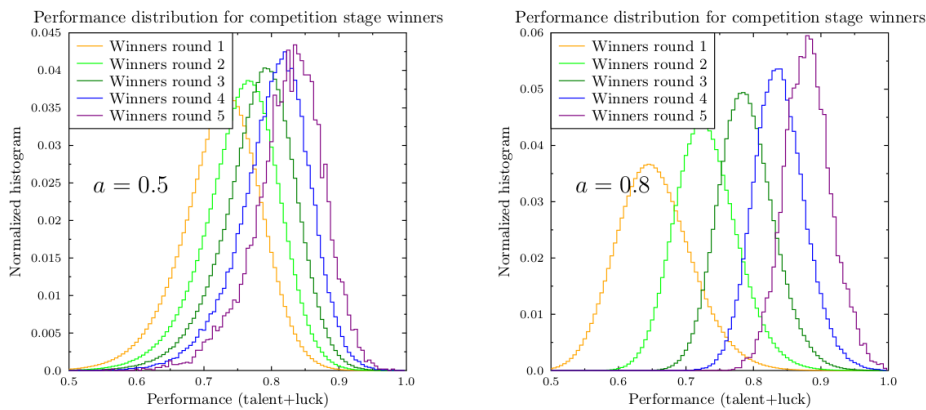

Figure 5: Performance distribution among the winners of five competition stages for narrow talent distribution ($\sigma_T = 0.1$). Left panel corresponds to medium value of the importance of talent ($a = 0.5$), while the right panel corresponds to relatively large $a = 0.8$.

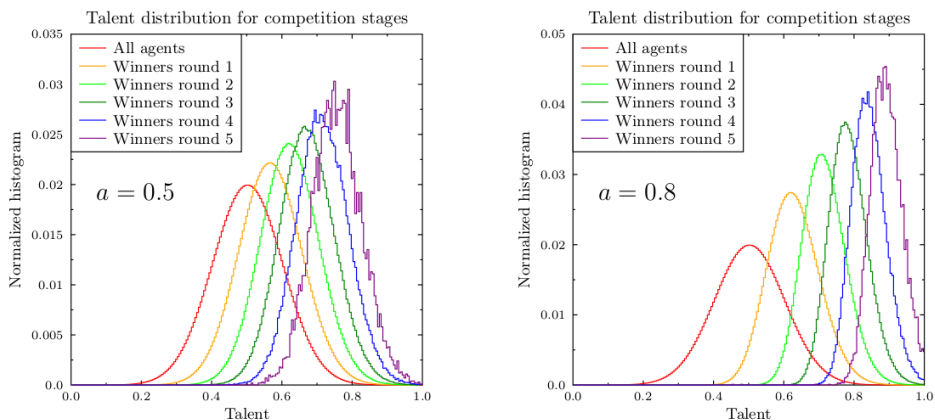

Figure 6: Talent distribution among the winners of competition stages 1–5, for narrow talent distribution ($\sigma_T = 0.1$). The parameter values are the same as in Figure 4. Increasing the talent contribution $a$ shifts the distributions of talent among the winners of consecutive stages towards higher values, but because there are very few agents with high talent values the distribution shape at the high end is determined by the number of available agents.

Summarizing the above results, we may state that in repetitive eliminations, talent counts. However, there is still a question whether the winners have to be 'lucky all the time'? We shall turn to this issue on the next section.

## 3.3 Are 'lucky streaks' important?

The model assumes that the luck distribution is a uniform one, between 0 and 1. Thus to be considered 'moderately lucky', an agent should have the luck value better than the median value of 0.5. Or we may pose a higher requirement, defining a very lucky agent as the one who has a luck greater than 0.75. Let us now ask the question: what is the distribution of the number of times that successful agents (for example the winners of rounds 5 and 6) have been moderately lucky or very lucky? The answer, for $\sigma_T = 0.2$ and $a = 0.5$ is in Table 0. As we can see, **there are almost no final winners who have not been consistently 'moderately lucky' almost all the time. Majority of them have been 'very lucky' in more than half of the competition rounds**. So, the statement that to be a winner one has to be blessed by fortune along entire life is justified. The difference in our large-scale, multistage tournament simulations and the ones of Pluchino et al. (2018) is that, in this case, being successful requires **both** high talent and a lot of luck. The competition for the few top places (single final winner or just ten finalists) is mainly among the most talented agents, for whom luck matters greatly. There is only so little room at the top... The final winner may not be the most talented agent (which would be chosen in a perfect, single comparison of talents without any luck component), but we may be quite sure that it is characterized by very high talent value.

On the other hand, we have to remember that the 'unlucky losers', at least those who made it to the final stages, are almost as talented, sometimes more talented – yet without the rewards associated with the success.

## 4. Empirical data

### 4.1 What type of data are we looking for?

Repeated comparisons of performance are present in many domains of social life. We have already mentioned exams at the transition points in the education ladder, but some forms of repeated comparisons with co-workers over work achievements are ubiquitous in the corporate – and even academic – life. These 'slow motion tournaments' certainly play a very important part in the overall success in life, and are very worthy of a detailed study. Unfortunately, it is very hard to choose appropriate quantitative measures or even to define a clear-cut set of conditions allowing meaningful comparisons of different social situations.

Sport competitions offer a chance of providing reliable quantitative data, in an environment in which fixed competition rules simplify the choice of the relevant factors. In our work we chose to analyse the iconic athletic competition, namely the 100 meter dash. There are several reasons for this choice. First, in contrast to many other sports disciplines, the 100 meter sprint involves a direct competition of a limited number of participants in a single race (typically less than 8). This forces the organizers of large events (such as the Olympic Games) to structure the whole competition in a tournament style similar to the one used in our model (in contrast to other disciplines, such as the Olympic marathon race, which involves all participants in a single competition instance). For most of the Olympic Games since 1968 there were four rounds for the 100 meter dash: Round One (RO), Quarter Finals (QF), Semi Finals (SF) and the Final (FIN) run. In the latest two games (London and Rio, 2012 and 2016) the Round One races were merged with the Quarter Finals. In contrast to our model, where there is only one winner in a group of ten, in Olympic Games typically three or four best runners in a given race qualify for the next round. This difference leads to a slower 'attrition rate' in the Olympic Games than in our model, which could have an influence on the results.

The second reason for the choice of the 100 meter sprint is its simple set up and ability to consistently compare results of many events: the track is the same in all competitions, the results are measured mechanically, without human judgement, the provisions for wind speed lead to an exclusion of results obtained in 'too favourable' conditions.

Lastly, while there has been some progress in the absolute speeds in the studied time (since 1968) it has been moderate.

## 4.2 Analysis of 100 meter dash performance in Olympic Games

To make the transition from the recorded run times to the abstract notion of performance we have used a simple rescaling. Performance is calculated as the ratio of the world record time $t_{WR}$ at the time of the race to the individual run time $t$, so that $p = t_{WR}/t$. In such approach, the runner beating the world record has the 'perfect' performance $p = 1$. Anyone running slower than the world record has $p < 1$. Moreover, this choice corresponds reasonably well to the choice of the centre of the talent distribution placed at 0.5. This value of the performance would correspond to the time of roughly 20-22 seconds for all the agents, depending on the normalization. It is a rather reasonable estimate of the 'average man' time for the 100 meter sprint (fastest times for amateurs are at the 13–14 second range, while the typical jogging speed would lead to about 25–30 seconds for a 100 meter distance). This estimate is further supported by extrapolating the data gathered for healthy young males by Alfano et al. (2017).

The rescaling has the additional advantage of making the performance of men and women comparable, although, as we shall show, not identical.

Figure 7 presents the time evolution of the performance averaged over the competition rounds in thirteen Olympic Games since 1968, for both men and women. There is significant variability of the results between the Games. Moreover, the discontinuation of the four rounds policy in 2012 and 2016 (and the associated increase in the number of the number of participants and heats in Quarter Finals) has led to a significant decrease of the average result in the Quarter Finals. For this reason it seems sensible to combine the results of Round One and Quarter Finals into single averages.

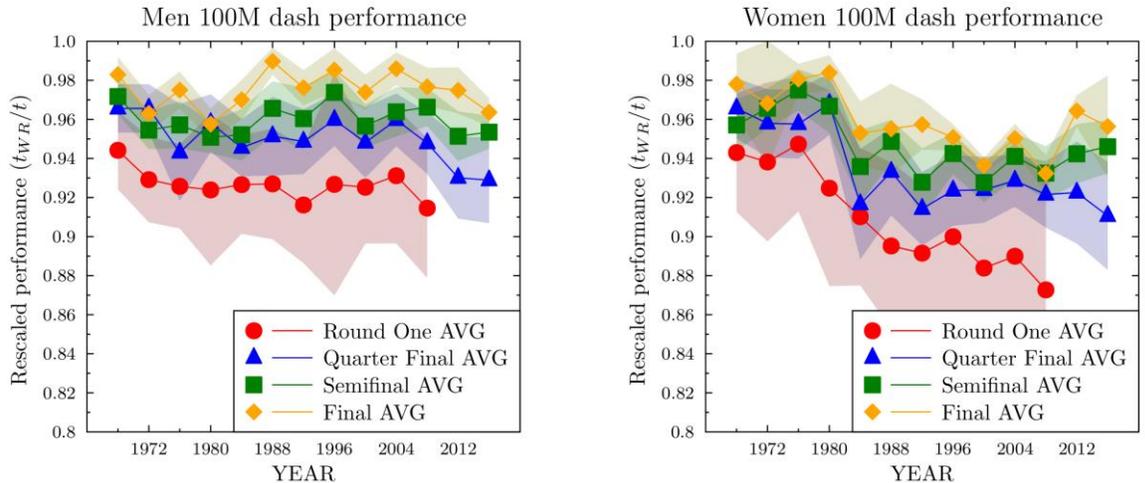

Figure 7: Average performance for competition rounds for the Olympic Games since 1968. Lines and dots denote the average values, while colour bands correspond to the standard deviations of the individual results at each stage of the Olympic Games event. Data source: SportsReference WEB page (2018).

The performance values averaged over all the Olympic Games studied (combining the RO and QF results) are presented in Table 2.

Of course, these averages provide only partial information about the results of the Olympic Games. A more detailed view is provided by the distribution of the performance at each competition stage. Figure 8 presents the histograms of such distributions for the three stages (RO+QF, SF and FIN) for all the Games in the data set.

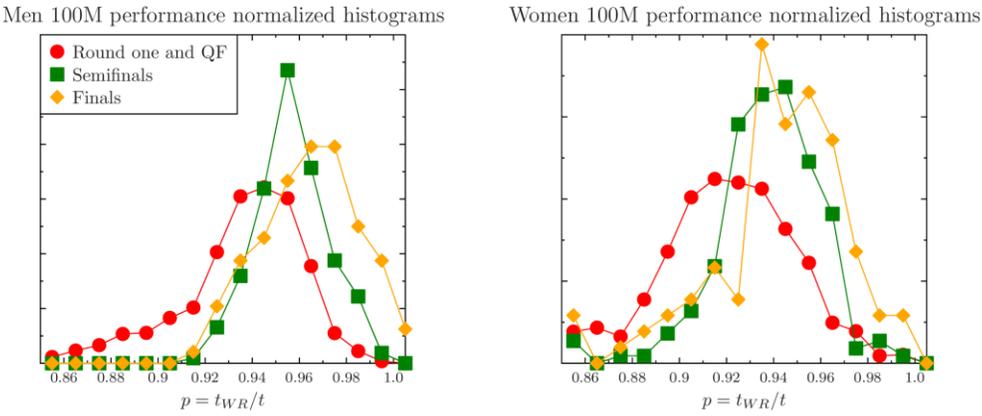

Figure 8: Histograms of the distribution of individual performances at each of the three stages of the Olympic competitions (RO and QF combined, SF and FIN). Data source: SportsReference WEB page (2018).

The fact that some of the participants have to run multiple times during the same Games allows us to look also into the individual variability of performance. Figure 9 presents the histogram of the performance differences in subsequent runs at the same event for all the participants who have run at least two times. A positive value of the performance difference means that the subsequent run was faster than the previous one, negative – that it was slower.

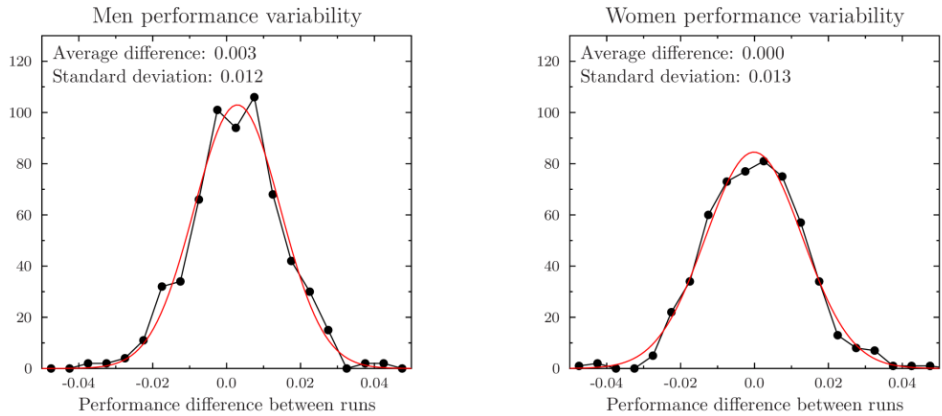

Figure 9: Individual performance variability, described as a histogram of the performance differences between subsequent runs of the same participant in the same Games. Black lines and dots are the averages for all the Games in the data set. Red lines are fits of a Gaussian distribution. The positive value of the average for men is highly statistically significant ($p < 0.0001$).

In general, the distributions of the differences in individual times between subsequent races by the same participant (or in rescaled performance) are very nearly Gaussian. The performance variability, as measured by the standard deviation is similar for men and women, around 0.012–0.013. These numbers are very similar to the performance variability data for other elite sports, especially similar high power, short duration disciplines such as swimming or cycling (Hopkins, 2005; Malcata and Hopkins, 2014; Paton and Hopkins, 2006; Pyne et al., 2004).

Interestingly, while the variability is similar for men and women, the distribution of the differences for women is centred at zero, which means that elite women runners are equally likely to run faster or slower in consecutive competition rounds. In contrast, the distribution for men is shifted slightly but statistically strongly significantly to positive values, which means that men tend to run faster in the later elimination stages. Such tendency is easily understood through folk psychology: the best participants run in earlier stages only as fast as needed to secure qualification to the next round, saving energy for the more demanding competitions at later stages, during which everyone would run faster. While plausible, such explanation begs the question: why such behaviour is not observed for women? Are they less 'calculating persons' and put more effort into each try?

## 4.3 Analysis of 100 meter dash performance in National Championships

To close this section, let us observe that the Olympic Games are just the pinnacle of the athletic competitions. In fact, typically, less than a hundred contestants compete there. They are the best that various national teams can send to the Games, moreover they have to fulfil some minimum performance criterion. Also, the number of participants from a given country can not be inflated too much. It is could be therefore interesting to look also at the competitive stages earlier than the Olympic Games.

In particular, we could look at the events leading to the choice of the participants in the Olympics at the national level. While the process of the candidate selection is not as straightforward as the Olympic tournament, we can get an estimate of the potential of the winners of the stage immediately below the Games if we consider the performance of the national champions. Figure 10 summarizes such data for 70 countries in the period between 1980 and 2006 (Data Source: GBRAthletics WEB page (2018)). Average performance values of $0.938 \pm 0.017$ for men and $0.898 \pm 0.027$ for women have been obtained and reported in the corresponding panels as red horizontal lines, surrounded by colored areas representing the standard deviations.

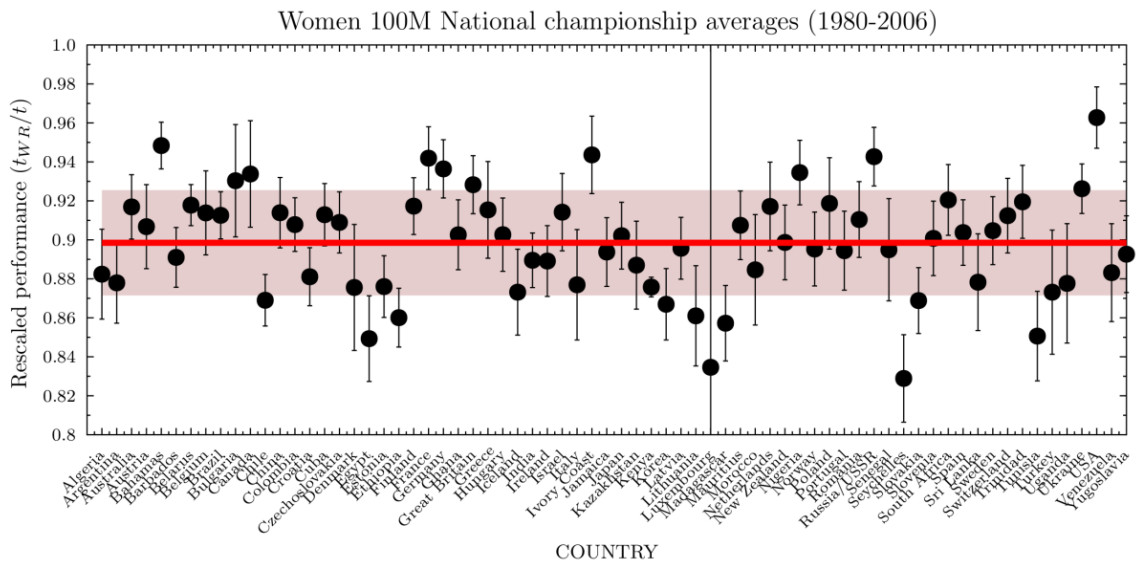

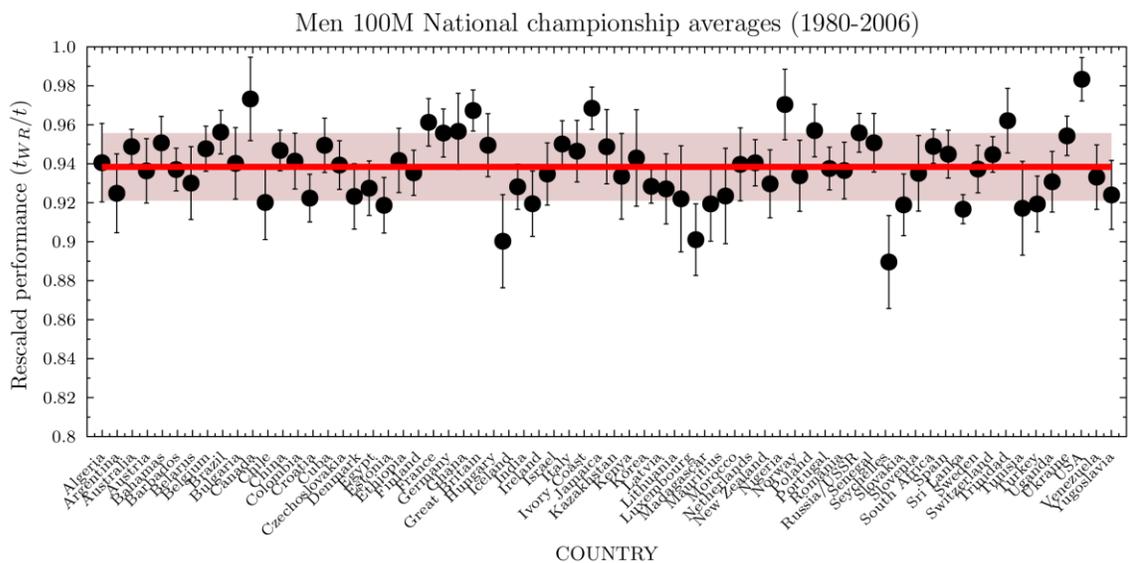

Figure 10: Average performance of national champions of the 100 meter dash in 70 countries for the period 1980–2006. Error bars denote standard deviation for each country. Red line is the overall world average, while the red shaded band denotes the standard deviation for the whole data set. (Data Source: GBRAthletics WEB page (2018)).

In the next section we will test the ability of our model in reproducing the average performance of athletes in both Olympic Games and National Championships. The further comparison among the performance distributions obtained with simulations and those observed in the last stages of Olympic Games will also allow to evaluate the relative contribution of both talent and luck in real competitions.

## 5. Comparing the model results with the sports data

As already noticed, there are numerous differences in the details of the actual 100 meter dash competitions and the assumptions of the model presented in Section 3. Nevertheless, despite its simplicity, we are going to show that the model allows to reproduce the empirical data remarkably well – and to provide an estimate of the role of luck at the highest levels of the iconic athletics competition, where only highly talented and trained people participate.

As noted earlier, our basic model has only two parameters: the width $\sigma_T$ of the talent distribution (since one can describe the uniform distribution by a Gaussian with $\sigma_T = \infty$) and the importance $a$ of the talent in obtaining a certain performance. We have already shown that their effects on the performance distributions of the winners of rounds 3, 4 and 5 are rather different. On one hand, decreasing $\sigma_T$ (i.e. making the talent distribution narrower and the number of highly talented agents smaller) shifts the performance averages to smaller values, especially for earlier competition rounds (as there might not be enough talented participants) and makes the distributions of performances for these rounds broader. On the other hand, decreasing $a$ broadens the performance distribution in a characteristic 'flat-top' shape (as the luck component acts in an 'equalizing' way). In this section we will show that it is possible to select the best-fit values of these parameters in order to match the real data for both the Olympic and the National championships results.

### 5.1 Comparison with Olympic results

In order to address the comparison with Olympic data, it is natural to identify the three Olympic stages (RO+QF, SF and FIN) with the three last rounds in the model (rounds 4–6). We draw the attention to the fact that, in analogy with real data, we will consider here the performance distributions of simulated *participants* at stages 4, 5 and 6, not to be confused with what has been done in the previous sections, where the performance distributions of the *winners* at rounds 3, 4 and 5 were presented. Of course, winners at a given round do coincide with participants at the subsequent round, but in Section 4 the simulated performance distributions of those agents were calculated at different stages (the previous ones) with respect to what we will do now.

Table 1 presents the averages of the performance in model simulations, obtained using best-fit parameters for participants of stages 4–6. The best choice, for both men and women, is to adopt quite narrow Gaussian distributions of talent, with $\sigma_T$ equal to 0.15 and 0.14 respectively, and very high values for the talent parameter $a$, 0.96 and 0.94 respectively. With this setting, a very good correspondence with the Olympic Games data is observed, as can be verified by a comparison with Table 2.

These best-fit values for the model parameters are able to reproduce not only the real average performances at each competition stage, but also the individual performance distributions for each stage, as shown in Figure 11 for men (left panel) and woman (right panel). It is evident that the obtained distributions closely resemble the empirical data shown in the corresponding panels of Figure 8.

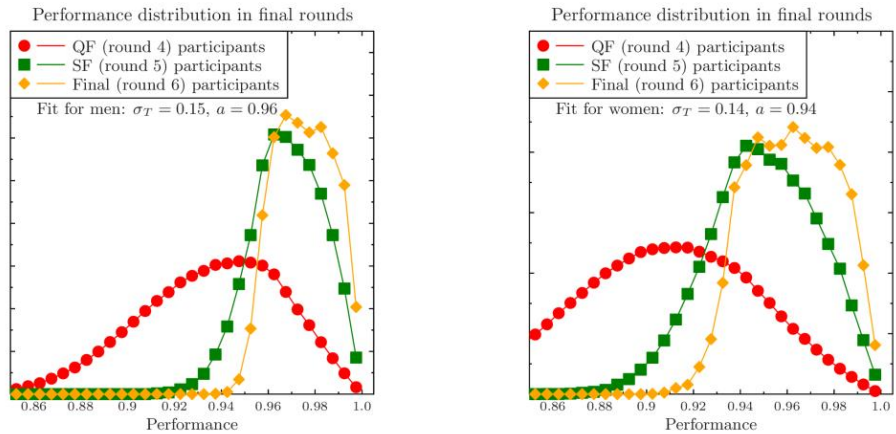

Figure 11: Simulated performance distributions of participants at the last three tournament stages (4–6), corresponding to the Olympic QF, SF and final races. Compare with Figure **Error! Reference source not found.**.

## 5.2 Comparison with National championship results

To compare the simulations with the national championships data we have to take a decision about how to identify the national championships in the model tournament structure. As we have identified the Olympics stages RO+QF, SF and FIN with rounds 4–6 of the simulations, it seems natural to identify the national championships as the stage immediately preceding the Olympics, i.e corresponding to round 3 in the simulations. As the data for the national events covers the winners, we can compare the empirical values with the averages for the winners of round 3 of simulations. Using the same parameters as the ones that gave the best fit for the Olympic Games (i.e. $\sigma_T = 0.15$, $a = 0.96$ for men and $\sigma_T = 0.14$, $a = 0.94$ for women) results in the average performance of the *winners* at stage 3 equal to 0.939 for men and 0.915 for women. This compares reasonably well with the previously found empirical values of $0.938 \pm 0.017$ for men and $0.898 \pm 0.027$ for women (see Figure 9). The higher discrepancy for women may be traced to a much higher disparity in national championship results. Quite a few countries share low performance with very small sizes of the community of women active in the discipline. This may be due to the country size or cultural heritage[4].

## 5.3 Discussion: role of luck in 100 meter dash

The high best-fit values of $a$ found for both men and women mean that, at least at these last stages of the competition, luck plays a minor role. On the other hand, this role could probably be also crucial. In fact, looking in Figure 12 at the talent distributions of participants to the last three simulated rounds (the same agents whose performance distributions have been successfully compared with real data in Figure 10), we notice a certain overlap among the various curves. Extrapolating this simulation result to the real world, this would mean that, even if a high talent is strictly necessary to reach the final stages of the competition, it can be not sufficient since we could find, in the final race, athletes less talented than others eliminated at previous rounds: in this context, small random events (good luck or bad luck) could make the difference.

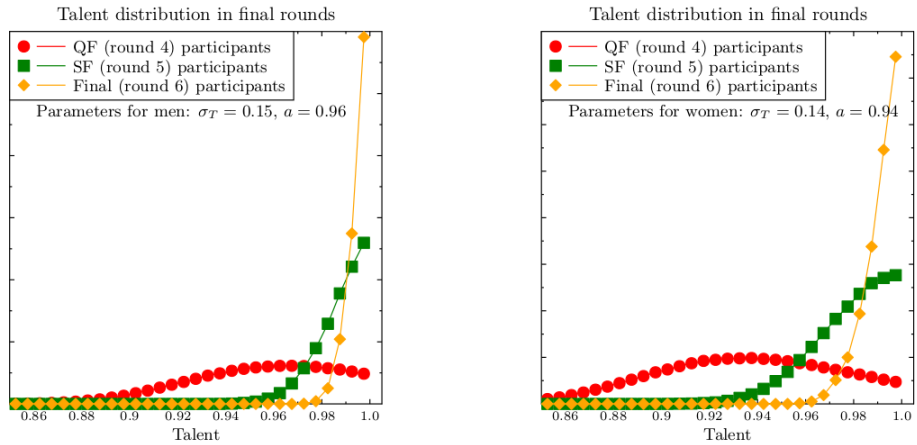

Figure 12: Simulated talent distributions of participants at the last three tournament stages (4–6), corresponding to the Olympic QF, SF and final races. Parameters are the same as in Figure 11. The talent of semifinalists and finalists has much narrow distribution as performance, because at this skill levels, the chance plays mainly the role of bad luck. Thus all finalists are extremely talented – but there can be only one winner, and only three places on the podium.

There is another way of estimating the luck role: by using the individual variability data. As the individual talent remains unchanged in our model, and the weighted luck component is a uniform distribution between 0 and $(1 - a)$, the difference in performance in subsequent runs has a symmetrical, triangular distribution shape limited between $-(1 - a)$ and $(1 - a)$. The standard deviation for such distribution is given by $(1 - a)/\sqrt{6}$. This would lead to the estimate of the individual variability of $0.04/2.45 = 0.016$ for men, and $0.06/2.45 = 0.024$ for women. Again, these values are remarkably close to the standard deviation of 0.012 and 0.013 obtained from the empirical data.

---

[4] If we exclude only 6 countries with the lowest averages (Seychelles, Luxembourg, Egypt, Tunisia, Madagascar and Ethiopia), the average performance jumps to 0.905. These countries have relatively small groups of sactive sportswomen, due either to cultural heritage or size. The exclusion of the corresponding data results in a significantly better fit between the ABM model and the empirical data.

## 5.4 Perceived injustice in tournament system

From the last considerations it follows that, particularly for the studied case of the tournament competition in sports, a highly asymmetric differentiation of rewards among different levels near the top of the competitive hierarchy could easily be interpreted as an unfair mechanism, differentiating athletes with very similar skills. Indeed, after several stages of a tournament competition, it is natural that the remaining participants – who are competing for the final victory – have very narrow differences in abilities and effort put into their preparation. Therefore, both simulated and real data suggest a strong sense of injustice in the relation between reward differences and the differences in the talent of the agents involved.

This injustice is typically associated with exponential rewards (such as the Wimbledon ones), or fees paid for the 'invited' participation of world and Olympic champions in smaller events, designed to boost the status and income of these events. We argue here that such sense of injustice would be present even if the reward system would not be exponential, but linear in relation to the achieved rank.

Let's consider the groups of agents who finish the competition at stages $t = 1,\ldots,6$, losing at this stage. These are the agents who lost the round $t$ competition. We can add the ultimate winner by assigning it the index $t = 7$. We shall denote the average talent in each group by $T_{Lt}$ (where $L$ stands for 'losers'). We define the difference in payoffs in the linear model by $Q$ and for simplicity set $Q = 1$, in whatever is the appropriate monetary unit. We can now introduce the 'unjust payoff disadvantage' $Z_t$, defined as $Z_t = Q/\Delta T_{Lt}$, where $\Delta T_{Lt} = T_{L(t+1)} - T_{Lt}$. The idea of such measure is the following: if the average talent of group $i$ is much smaller than of the next group ($t = 1$), then $\Delta T_{Lt}$ is large and $Z_t$ is small. But if the talent difference between the two groups is small, then the injustice defined by $Z_t$ would be large, making it a natural and simple measure of the phenomenon. For consistency of presentation we additionally normalize $Z_t$ so that $Z_1 \equiv 1$.

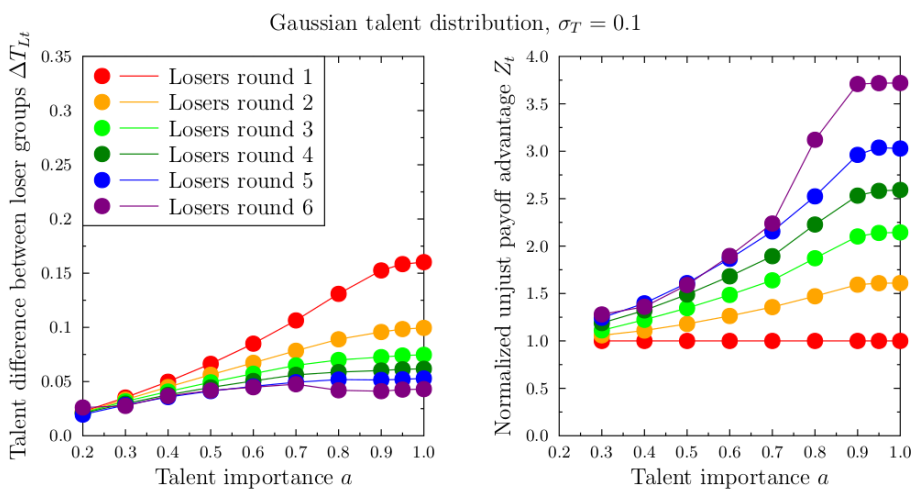

Figure 13: Average talent difference $\Delta T_{Lt}$ (left panel) and the normalized unjust payoff disadvantage $Z_t$ (right panel). Both are presented as functions of the role of talent in the tournament *a*, for fixed $\sigma_T$=0.1.

Figure 13 presents the dependence of the average talent difference $\Delta T_{Lt}$ (left panel) and the normalized unjust payoff difference $Z_t$ (where the normalization results in $Z_1 = 1$, right panel) on the role of talent in the tournament $a$, for $\sigma_T = 0.1$. The talent differences become smaller at the later stages of the competition, making the perceived injustice, measures by $Z_t$, greater. The effect is even stronger in the case when there are many agents with talent near the top value, for example when $\sigma_T = 0.2$. Figure 14 presents the results for such case. The drop in the talent differences is more pronounced, making the perceived injustice enormous.

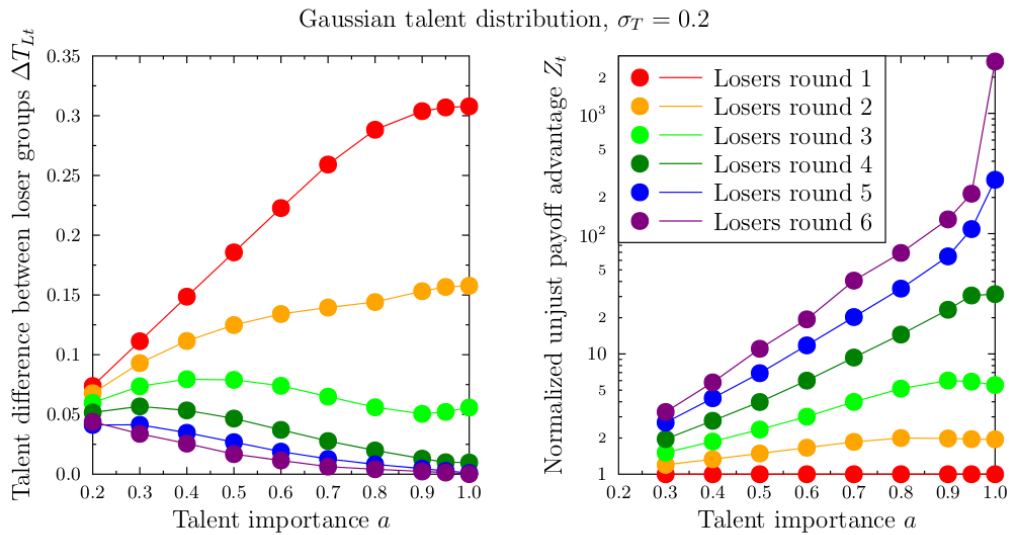

Figure 14: Average talent difference $\Delta T_{Lt}$ (left panel) and the normalized unjust payoff disadvantage $Z_t$. The normalization results in $Z_1 = 1$. Simulations assumed $\sigma_T = 0.2$. Note the logarithmic scale on the right panel: due to very small differences in the average talent at late competition stages, the values of $Z_t$ become very large. In the case when luck plays no role ($a = 1$) the normalized unjust payoff disadvantage reaches 2700 for the finalists, comparing themselves to the ultimate winner (i.e. it is 2700 times greater than the disadvantage perceived by the agents at the bottom of the hierarchy).

Despite the arbitrary nature of the choice of $Z_t$ as the quantity measuring the relative injustice, it clearly shows that even relatively 'conservative', linear ways of increasing the payoffs at higher levels of the hierarchy result in differences in rewards disproportional to the corresponding differences in talent and effort. Moreover, when the advancement to the higher levels depends significantly on random luck, these differences would be perceived as highly unjust. Whatever way is used to compare talent (merit) and reward, if the talents are similar and rewards different, one can expect a feeling of injustice.

Contrary to the standard sociological perception, this injustice is not concentrated among the agents at the bottom of the hierarchy (who have relatively low values of talent) but among the most talented 'final losers', who were stuck at the later stages of the competition. Not only are they almost as talented as the highly rewarded winners, they have also 'invested' the same effort and perseverance to attain the final round ... only to see the winner 'taking it all'. We note that in presenting $Z_t$ we have used the linear payoff growth; adopting a 'Wimbledon-like' exponential one, these perceived disadvantages would be much greater.

## 6. Conclusions

In conclusion, using a relatively simple model, we were able to estimate the role of chance in the Olympic level 100 meter events. This estimation is supported by the comparisons of the model results in both group and individual variability of performance, as shown in the previous section. Random luck accounts for about 4% of performance for men, and 6% for women. This is an astonishingly large value, considering the efforts taken to diminish the role of external conditions in the competition and the fact that at the Olympic level, it involves a relatively small number of elite athletes. The value of 4% could therefore act as as a strong lower limit in estimates of the role of chance in other, less controlled situations.

Recognizing the presence of chance, and its non-negligible influence even in a situation as tightly controlled and standardized as the 100 meter sprint, may serve as a reminder to the beneficiaries of any comparison based rankings that their success is most likely, to some extent, due to random luck. They should realize that, most likely, there are some people below their rank more 'deserving' of a win (in terms of broadly understood talent and effort). Moreover, the widening gap between the rewards given to winners and to 'almost winners' – which in the winner-takes-all context naively overlaps the distinction between winners and losers – is even less justified than commonly understood.

The potential offered by agent-based approaches to go beyond 'anecdotal' evidence and the capacity to provide flexible and quantitative understanding of the modelled phenomena is increasingly recognized in social sciences. It is further strengthened by coupling these approaches with analyses of data from social networks and other Big Data sources. Since inequalities are so ubiquitous in our societies, the insight offered by models may be quite useful in shaping socially responsible policies and understanding – hopefully, predicting – the policy outcomes.

## List of abbreviations

RO – Round One (of Olympic 100 meter sprint competition);
QF – Quarter Final;
SF – Semi Final;
FIN – final run.

of films: How much is a movie star worth? *Journal of cultural economics*, 17(1):1–27.

Wikipedia WEB page a (2018). List of highest paid film actors. en.wikipedia.org/wiki/List_of_highest_paid_film_actors Accessed: 26. Dec 2018.

Wikipedia WEB page b (2018). List of most-liked facebook pages. en.wikipedia.org/wiki/List_of_most-liked_Facebook_pages Accessed: 26. Dec 2018.

Wimbledon WEB page (2018). Wimbledon prize money history. www.wimbledon.com/en_GB/aboutwimbledon/prize_money_and_finance.html Accessed: 26. Dec 2018.

Wittig, M. A., Marks, G., and Jones, G. A. (1981). Luck versus effort attributions: Effect on reward allocations to self and other. *Personality and Social Psychology Bulletin*, 7(1):71–78.

Yucesoy, B. and Barabási, A.-L. (2016). Untangling performance from success. *EPJ Data Science*, 5(1):17.
## Tables

**Table 1**: Luck distribution among the winners of rounds 5 and 6 for simulations with $\sigma_T = 0.2$ and $a = 0.5$. Notice that there are 10000 winners of round 5 and 1000 winners of the final round 6 in the 1000 simulation runs.

| Number of times an agent has been: | moderately lucky $(L_j(t) > 0.5)$ | | very lucky $(L_j(t) > 0.75)$ | |
|---|---|---|---|---|
| | Round 5 winners | Round 6 winners | Round 5 winners | Round 6 winners |
| | 0 | 0 | 50 | 1 |
| | 1 | 0 | 519 | 7 |
| | 39 | 0 | 2201 | 76 |
| | 1249 | 4 | 4405 | 252 |
| | 8711 | 144 | 2818 | 432 |
| | — | 852 | — | 232 |

**Table 2**: Average performance at three competition stages: RO+QF, SF and FIN, calculated for the whole data set of thirteen Olympic Games. Compare with Table 2. Notice that the results concern participants at stage $t$, and not the winners at the previous stage $t - 1$.

| Competition round | Men | Women |
|---|---|---|
| Parameter values for best fit | $\sigma_T = 0.15$, $a = 0.96$ | $\sigma_T = 0.14$, $a = 0.94$ |
| | Average performance | Average performance |
| Model stage 4 (RO+QF) | 0.935 | 0.908 |
| Model stage 5 (SF) | 0.968 | 0.949 |
| Model stage 6 (Final) | 0.975 | 0.962 |

**Table 3**: Average performance at three competition stages: RO+QF, SF and FIN, calculated for the whole dataset of thirteen Olympic Games.

| Competition round | Men | | Women | |
|---|---|---|---|---|
| | Average performance | Standard Deviation | Average performance | Standard Deviation |

| Round One combined with Quarter finals | 0.934 | 0.032 | 0.915 | 0.046 |
| --- | --- | --- | --- | --- |
| Semi Finals | 0.959 | 0.014 | 0.947 | 0.020 |
| Final | 0.975 | 0.013 | 0.959 | 0.022 |